\documentclass[aps,showpacs,amsmath,amssymb,twocolumn,prb,floatfix,superscriptaddress,a4]{revtex4}
\usepackage{longtable}
\usepackage{graphicx}
\usepackage{epsfig}
\usepackage{amsmath}
\usepackage{amsfonts}
\usepackage{amssymb}

\begin{document}
\title{Performance of the coupled cluster singles and doubles method
  on two-dimensional quantum dots.}

\author{E. Waltersson}
\affiliation{Fysikum, Stockholm University, AlbaNova, S-106
91 Stockholm, Sweden}
    
\author{E. Lindroth}
\affiliation{Fysikum, Stockholm University, AlbaNova, S-106
91 Stockholm, Sweden}

\date{\today}

\begin{abstract}
An implementation of the coupled-cluster single- and double
excitations (CCSD) method on two-dimensional quantum dots is
presented. Advantages and limitations are studied through comparison
with other high accuracy approaches for two to eight confined
electrons. The possibility to effectively use a very large basis set
is found to be an important advantage compared to full configuration
interaction implementations. For the two to eight electron ground
states, with a confinement strength close to what is used in
experiments, the error in the energy introduced by truncating triple
excitations and beyond is shown to be on the same level or less than
the differences in energy given by two different Quantum Monte Carlo
methods. Convergence of the iterative solution of the coupled cluster
equations is, for some cases, found for surprisingly weak confinement
strengths even when starting from a non-interacting basis. The limit
where the missing triple and higher excitations become relevant is
investigated through comparison with full Configuration Interaction
results.
\end{abstract}

\maketitle
\section{Introduction\label{sec:intro}}

Ever since Tarucha {\em et al.}\cite{Tarucha} experimentally
demonstrated atom-like properties in few electron quantum dots, in
particular the existence of a shell structure, these systems have
attracted a lot of theoretical interest as new targets for many-body
methods.  In contrast to the situation for the naturally occurring
many-body systems, the strength of the overall confinement relative to
that of the inter-particle interaction can here be freely varied over
a large range, at least in theory, and thus completely new regimes can
be explored.  When the aim is to study the performance of a specific
many-body method, it is justified to use a simple model for the
confining potential and most theoretical studies restrict themselves
%
 to the two dimensional harmonic oscillator potential. The interaction
 between the dot electrons and the surrounding semi-conductor material
 is further usually modeled through the use of a material specific
 effective electron mass and relative dielectric constant.  This model
 implies thus a two-dimensional truly atom-like device, on which
 calculational methods developed for atoms can be applied after minor
 adjustments. Early calculations proved the model to be adequate.
 Combined with a reasonable account for the electron-electron
 interaction through methods such as Density Functional Theory
 (DFT)\cite{matagne02,melnikov05,koskinen97,macucci97,Lee_98} or
 Hartree-Fock\cite{Fuji_96, Bedn_99, yannouleas99,Giov_08}, the
 two-dimensional harmonic oscillator confining potential did indeed
 give a good qualitative agreement with experiments. In particular the
 closed shells forming with two, six and twelve trapped electrons
 could be explained, as shown in many studies, see e.g. the review by
 Reimann and Manninen\cite{reimannRMP02}.

Neither the true form of the dot confinement, nor the extent to which
it deviates from being purely two dimensional, is easily extracted
from experiments. If the electron-electron interaction could be
sufficiently well accounted for, it might be possible to gain further
understanding of such properties through comparison between experiment
and theory. Some studies in this direction have been performed,
e.g. by Matagne {\em et al.}\cite{MatagneRealisticVQD} who made
quantitative statements about the non-harmonicity of the confining
potential from the comparison between DFT calculations and
experiments.  DFT has proven to be able to account for a large part of
the electron correlation, but still it is not really the best choice
for such investigations since it is hard to make an a priori estimate
of the obtained error.
There are several ways to account for correlation more
systematically. A number of studies on quantum dots and related
structures have been carried out with Configuration Interaction (CI),
see e.g. Refs.~\cite{Bruc_00, Szaf_03,
  ReimannCI,Mikhailov:QDlithium,Mikhailov:QDberyllium,
  Rontani2006,popsueva,Waltersson_ring_09,kvaal:09,Saelen_ring_10}. Full
CI is in principle exact and applicable for all relative interaction
strengths. The term ``full CI'' refers to a calculation where all
Slater determinants, obtained by exciting all possible electrons to
all possible orbitals that are unoccupied in the studied electronic
configuration, are included. It is obvious that the size of the full
CI problem grows extremely fast both with the number of particles and
with the size of the basis set (used to represent the unoccupied
orbitals). It is well known that truncated CI, i.e. keeping e.g. only
single and double excitations from the leading configuration, lacks
size-extensivity. In short this implies that the method does not scale
properly with the size of the studied system, see for example the
review by Bartlett\cite{bartlett:81:review}. Truncation of the number
of excitations is thus not a real option, and instead all but the
smallest systems have to be calculated with very small basis
sets. Recently a thorough investigation of the performance of full CI
applied to quantum dots with a basis set consisting of harmonic
oscillator eigenfunctions was made by Kvaal\cite{kvaal:09} and the
main conclusion was that the convergence with respect to the size of
the basis set was slow and that additional features such as effective
two-body interactions have to be added for meaningful comparisons with
experiments.
An alternative approach for high accuracy calculations are Quantum
Monte Carlo (QMC) methods, which successfully have been applied to
quantum dots\cite{Saarikoski,Egger:99,Egger:err:99,
  montecarlo:00,Will_02,montecarlo:err:03,Ghosal,Weis_05,Zeng_09}. Here
the computational cost grows modestly with the number of electrons,
and the method provides a very efficient way to calculate the ground
state for a specific symmetry.  The nodal structure of the trial wave
function can be used to impose restrictions on the solutions so that
also excited states can be obtained to some extent, see e.g. the
discussion in the review by Foulkes {\em et
  al.}\cite{qmc:review:01}. Still, calculations on general excited
states are not straightforward and additional methods are needed for
realistic calculations on important parts of the quantum dot spectrum.

Most of the methods used on atoms and molecules have also been applied
to quantum dots in several implementations. The least studied method
is however many-body perturbation theory which has been shown to be
very powerful for the calculation of atomic properties. Calculations
up to second order in the perturbation expansion have been made by
just a few authors\cite{Slogget,waltersson:07}, and equally few
coupled-cluster calculations have been
presented\cite{henderson:00,heidari:07}.  For small to medium sized
molecules, as well as atoms, the coupled-cluster(CC) method is known
to successfully combine feasibility with accuracy. The coupled-cluster
theory was introduced in 1960 by Coester and
K\"ummel\cite{coester:60:short} in nuclear physics, and since then
contributions have been given by many authors.  A rather recent review
regarding its performance in quantum chemistry has been made by
Bartlett and Musia{\l}\cite{bartlett:rmp:07}. We present here a
thorough investigation of how the coupled-cluster method with single
and double excitations (CCSD) performs, in comparison with full CI and
Quantum Monte Carlo-studies, on two-dimensional quantum dots.  In
section~\ref{sec:method} we summarize CCSD and briefly discuss its
advantages. In section~\ref{sec:impl} our implementation for
calculations on circular quantum dots is outlined.  In
Section~\ref{sec:results} we present results for dots with two to
eight electrons and compare them with those obtained with other
methods. The first question is whether the restriction to single and
double excitations is adequate. It is known to be a good approximation
for atoms, but we expect it to eventually fail for sufficiently weak
confinement strengths. Here, we try to establish when this
happens. The next point is the feasibility and we show that results
converged with respect to e.g. basis size can be obtained for much
larger systems than is possible for CI calculations. We show for
$\hbar\omega \sim 3$~meV and the $N=2$ to $8$ groundstates that the
error relative to Diffusion Monte Carlo results is on the same level
or less than the difference between the energies given by the Variational and
Diffusion Quantum Monte Carlo methods.

\section{Theory\label{sec:method}}

The formalism used in the present study can be found in more detail in
the textbook by Lindgren and Morrison\cite{mbpt}. Here we just discuss
the aspects important for the understanding of the present results.
In order to solve the Schr\"{o}dinger equation
\begin{equation}
\label{sch}
H \Psi = E \Psi,
\end{equation}
for an $N$ -fermion system,
the Hamiltonian is partitioned  as
\begin{equation}
\label{perturb}
H  = H_0 +V,
\end{equation}
where the eigenstates of $H_0$ are known and $V$ is the
remainder, i.e. it is the perturbation with respect to the already
solved Hamiltonian $H_0$.  In the present study $H_0$ is usually the
Hamiltonian for the non-interacting system, and thus $V$ is the whole
electron-electron interaction, but also other choices have been
examined as will be discussed below.
We will further assume that $H_0$ is  a sum of $N$
 single-particle Hamiltonians with known eigenstates:
\begin{eqnarray}
\label{H0h}
H_0  & = & \sum_i^N h_i , \nonumber \\ 
h_i \mid i \rangle & = & \varepsilon_i  \mid i \rangle.
\end{eqnarray}
 With the orbitals, $\mid i \rangle$, we can form a {\em model space}
suitable for the state(s) we are interested in. In the following we
will use a one-dimensional model space, $P$, spanned by one Slater
determinant
\begin{eqnarray}
\alpha & = & \left\{a b c d \ldots N\right\}, \label{Pspace} \\ P & = & \mid \alpha\rangle \langle \alpha \mid,
\end{eqnarray}
where $a,b,c,d \ldots ,N$ denote the occupied orbitals, and the curly brackets denote antisymmetrization.  A
multi-dimensional {\em extended} model space is also a
possibility\cite{mbpt}, which though will be left for future investigations.  The
{\em model function} is the projection of the exact solution,
Eq.~(\ref{sch}), onto the model space
\begin{equation}
\Psi_0 = P \Psi.
\end{equation}
We further assume that the model function, but not the full wave
function, is normalized, a condition usually referred to as {\em
  intermediate normalization}.  It is possible to define a {\em wave
  operator}, $\Omega$ that transforms the model function into the
exact state, i.e.  $\Psi = \Omega \Psi_0$.  To separate the part
of $\Omega$ that projects onto the model space and that which brings
the solution out of the model space we write
 \begin{equation}
\label{omega}
\Omega= 1 + \chi,
\end{equation}
where $\chi$ is sometimes referred to as the correlation operator.
The wave operator can be obtained from the {\em
  generalized}\cite{lindgren:74:mbpt} {\em Bloch
  equation}\cite{bloch:58}, which we will use in the
form\cite{lindgren:74:mbpt,mbpt};
\begin{equation}
\label{eq:bloch}
 \left[ \Omega , H_0 \right] P = \left( QV \Omega P -\chi PV \Omega P\right),
\end{equation}
 where $Q$ is the orthogonal space to $P$, such that $P+Q = 1$.
 Eq.~(\ref{eq:bloch}) is equivalent to the Schr{\"o}dinger equation,
 but in this form it allows for an iterative solution
 procedure. Setting $\chi$ on the right hand side initially to zero,
 one can obtain a first approximation of $\chi$ which then can be
 inserted on the right-hand side to get a better approximation and so on
 until convergence is reached. When an order by order expansion is
 carried through, it can be shown that in each order the so called
 {\em unlinked} diagrams from the first term in Eq.~\ref{eq:bloch} are
 canceled by contributions from the second term. 
 Unlinked
 contributions are those that include parts that are surrounded by
 $P$-operators, as the$\chi PV \Omega P$-term in Eq.~\ref{eq:bloch}.
 This is the so called
 {\em linked diagram theorem}, see for example
 Refs.\cite{mbpt,bartlett:rmp:07} and references therein. It is thus
 possible to keep only linked contributions in the expansion.  While
 the exclusion principle is obeyed for the sum of linked and
 unlinked contributions, the cancellation of the unlinked
 contributions is achieved only if the exclusion principle is lifted,
 resulting in exclusion principle violating diagrams also in the
 retained linked contributions. It is the linked expansion which will
 be the basis for the coupled cluster expansion below.

When $\Omega$ has been obtained it can be used to construct the {\em
  effective Hamiltonian}
\begin{equation}
H_{\textrm{eff}}= PH_0P + P V \Omega P,
\end{equation}
  which gives the exact energies when acting on the model
  space~\cite{mbpt}. The total energy can then be written as
\begin{equation}
\label{deltaE}
 E = \langle \alpha | H_{\textrm{eff}}| \alpha \rangle = \langle \alpha | H_0  + V +  V \chi  |\alpha \rangle,
\end{equation}
where $P \mid \alpha \rangle = \mid \alpha \rangle$ has been used.

$Q$ is the complementary space to $P$ and 
can formally be built up from all Slater
determinants, $\beta$, that differ from $\alpha$ 
\begin{equation}
Q = \sum_{\beta \ne \alpha} |\beta\rangle \langle \beta|.
\end{equation}
The space spanned by $Q$ is in principle infinite, but in
practice we use a finite basis set to represent the eigenstates to
$h$, Eq.~(\ref{H0h}). This makes also the $Q$-space finite, but still it
grows rapidly both with the size of the basis set and with the number
of particles. 
We focus now on the representation of the $Q$-space for several particles.
For this purpose we can classify the Slater determinants belonging to
$Q$ with respect to by how many single particles states they differ
from $P$.  For example
\begin{equation}
\alpha_a^r= \left\{r b c d \ldots N\right\}, 
\label{single}
\end{equation}
differs from Eq.~(\ref{Pspace})  only in that $a$ has been replaced by $r$, and it is labeled a {\em single} excitation, while
\begin{equation}
\alpha_{ab}^{rs}= \left\{r s c d \ldots N\right\}, 
\label{double}
\end{equation}
differs from Eq.~(\ref{Pspace}) in that $a$ and $b$ have been replaced
by $r$ and $s$, and it is labeled a {\em double} excitation. A
complete calculation on a two-particle system requires single and
double excitations, while such a calculation on a three particle
system also requires triple excitations, and so on. For a general
many-particle system it is necessary to truncate this series at some
point due to both complexity and computational load.  For this
truncation there exists several choices. $\chi$ is the part of the
wave operator that lies in the $Q$-space. It can for example be
divided up as
\begin{equation}
\label{chi_truncation}
\chi =\chi_1 + \chi_2 + \chi_3 + \ldots,
\end{equation}
where the subscripts denote the number of excitations.  If we truncate
this sum after e.g. $\chi_3$, we will reproduce CI with single,
double, and triple excitations\cite{bartlett:rmp:07}.  With the
coupled-cluster approach the truncation is made in an alternative way.
First we start from the linked form of the Bloch equation
\begin{equation}
\label{eq:blochlinked}
 \left[ \Omega , H_0 \right] P = \left( QV \Omega P -\chi PV \Omega P\right)_{linked},
\end{equation}
where only linked contributions are retained in the iterative
procedure. As a second step we define a \emph{cluster operator} $S =
S_1 + S_2 + S_3 + \ldots$, where each term represents the {\em
  connected} part of the wave operator for $n$ excitations, $S_n =
\left(\Omega_n \right)_{connected}$.  The term connected denotes that
the wave operator cannot be divided up in parts where the particles
interact independently in smaller clusters, e.g. two-by-two.  The $S$
operator can be shown\cite{Lindgren:78:cc} to satisfy a Bloch type
equation
\begin{equation}
\label{eq:soperator}
 \left[ S, H_0 \right] P = \left( QV \Omega P -\chi PV \Omega P\right)_{\rm connected}.
\end{equation}

The wave operator, $\Omega$, can now be written as a sum of products
of $S_n$-operators. All such terms are generated through the {\em
  exponential ansatz}:
\begin{multline}
\Omega = \{ exp\left( S \right) \} = 1 +
S_1 + S_2 + \frac{1}{2!}\{S_1\}^2 + \{S_1 S_2\} + \frac{1}{3!}\{S_1^3\} \\
+ \frac{1}{2!}\{S_2^2\} + \frac{1}{2!}\{S_1^2 S_2\}+\frac{1}{4!}\{S_1^4\} + \ldots.
\end{multline}
The curly brackets denote here that it is the normal ordered products
of the operators that should be used, which implies
antisymmetrization.  We can now identify all single, double, triple
etc excitations accordingly
\begin{eqnarray}
\label{scluster}
\Omega_1 &=& \boxed{S_1} ,\nonumber \\
\Omega_2 &=& \boxed{S_2 + \frac{1}{2} \left\{ S_1^2 \right\}}, \nonumber \\ 
\Omega_3 &=& S_3 + \boxed{ \left\{ S_1 S_2 \right\} +  \frac{1}{3!}\left\{ S_1^3 \right\} },\nonumber \\
\Omega_4 &=& S_4 +  \left\{ S_1 S_3 \right\} + 
\boxed{ \frac{1}{2} \left\{ S_2^2 \right\} +  \frac{1}{2} \left\{ S_1^2 S_2\right\}
  + \frac{1}{4!} \left\{ S_1^4 \right\}}, \nonumber \\
 \Omega_5 &=& \ldots.
\end{eqnarray}
From Eq.~\ref{scluster} it is clear that when
truncating after the $S_2$ cluster, 
we  still include the parts of
$\Omega_3$ and $\Omega_4$ that can be written as combinations of $S_1$
and $S_2$ operators, i.e. the terms in boxes above. This is  the Couple Cluster Singles and Doubles method. 
How these products of $S$-operators enter in the expansion 
will become more clear when 
Eqs.~(\ref{S1}-\ref{S2}) are discussed below. See also Ref.\cite{sos:90:ccsd} for more
details.  There are two clear advantages of this truncation
scheme. First, the probably most important triple and quadruple
excitations are now included in a scheme that is much less
computationally demanding than calculating full triples and
quadruples. Second, and this is in contrast to the scheme indicated in
Eq.(\ref{chi_truncation}), the inclusion of the disconnected products
makes the coupled-cluster method size extensive also in its truncated
version. 

 In the following we will investigate the performance of the coupled
 cluster method when including all $S_1$ and $S_2$ terms in
 Eq.(\ref{scluster}) (the expressions in boxes), i.e. the CCSD
 method. Although the practical implementation is different, the
 present study includes the same effects as the implementation for
 atoms by Salomonson and {\"O}ster\cite{sos:90:ccsd}, where more
 details about the method can be found.

\section{Implementation\label{sec:impl}}
\subsection{Single-particle treatment}
For a single particle confined in a circularly symmetric potential 
the Hamiltonian reads
\begin{equation}
\label{hone}
\hat{h} =\frac{\mathbf{\hat{p}}^2}{2m^*} + \hat{u}_c(r),
\end{equation}
where the
effective electron mass is denoted with $m^*$.
With a pure harmonic confinement we have  
\begin{equation}
\label{harmpot}
\hat{u}_c(r)=\frac{1}{2}m^*\omega^2r^2.
\end{equation}
This is the confining potential used in all numerical results in the
present study but any circularly symmetric confinement can 
be used in the developed computer code.

The single particle wave functions separate in polar
coordinates as
\begin{equation}
\label{wavefunctionexpansion_eq}
\Psi_{nm_lm_s}(r,\phi) = u_{nm_lm_s}(r) e^{im_l\phi} |m_s\rangle.
\end{equation}

We expand the radial part of the wave functions in so
called B-splines labeled $B_i$ with coefficients $c_i$, i.e.
\begin{equation}
 u_{nm_lm_s}(r)  = \sum_{i=1} c_i B_i(r).
\label{Bspline_expansion_eq}
\end{equation}
B--splines are piecewise polynomials of a chosen order $k$, defined on
a so called knot sequence and they form a complete set for the linear
space defined by the knot sequence and the polynomial
order~\cite{deboor}.  Here we have used $40$ points in the knot
sequence, distributed by the use of an arcsin-function. The last knot
point, defining the boundary of the box to which we limit our problem,
is scaled with the potential strength through the harmonic oscillator
length unit $\sqrt{\hbar/(m^* \omega)}$. For example with $\hbar
\omega \approx 11.857$~meV (which corresponds $1$ a.u.$^*$ for GaAs, see
Section~\ref{sec:results}) the last knot point is located at
$r\sim 70$~nm. The polynomial order is $10$ and combined with the knot
sequence this yields $29$ radial basis functions, $u_{nm_lm_s}(r)$, for
each combination $(m_l,m_s)$. The lower energy basis functions are
physical states, while the higher ones are mainly determined by the
box. The unphysical higher energy states are, however, still essential
for the completeness of the basis set.

Eq.~(\ref{wavefunctionexpansion_eq}-\ref{Bspline_expansion_eq})
implies that the one-particle Schr\"odinger equation, (\ref{hone}), can be
written as a matrix equation
\begin{equation}
\mathbf{hc}=\epsilon\mathbf{Bc}
\label{Matrix_eq}
\end{equation}
where $h_{ij}=\langle B_i
e^{im\theta}|\hat{h}|B_j e^{im\theta}\rangle$ and
$B_{ij}=\langle B_i|B_j \rangle$
\footnote{Note that $\langle B_j|B_i \rangle\neq\delta_{ji}$ in general since B--splines of order
larger than one are non--orthogonal.}.

Eq.~(\ref{Matrix_eq}) is a generalized eigenvalue problem that can be
solved with standard numerical routines. The integrals in
(\ref{Matrix_eq}) are calculated with Gaussian quadrature. 
B-splines are piecewise polynomials, and since also the potential is in polynomial
form in  Eq. (\ref{hone}),   essentially no
numerical error is produced in the integration. 

\subsection{The Coulomb interaction}
The perturbation $V$ in Eq.~(\ref{perturb}) will include the
electron-electron interaction not accounted for in $H_0$. It can be
the full Coulomb interaction or the difference between that and some
mean-field approximation. In either case, we need a suitable way of
dealing with the Coulomb interaction in two dimensions. As suggested
by Cohl {\em et al.}\cite{Cohl}, the inverse radial distance can be
expanded in cylindrical coordinates $(R,\phi,z)$ as
\begin{equation}
\label{multipole_eq}
\frac{1}{|\mathbf{r}_1-\mathbf{r}_2|} =
\frac{1}{\pi\sqrt{R_1R_2}}\sum_{m=-\infty}^{\infty}Q_{m-\frac{1}{2}}(\chi)e^{im(\phi_1-\phi_2)},
\end{equation}
where
\begin{equation}
\label{chi_eq}
\chi = \frac{R_1^2+R_2^2+(z_1-z_2)^2}{2R_1R_2}.
\end{equation}
Assuming a two-dimensional confinement we set $z_1=z_2$ in
(\ref{chi_eq}).  The $Q_{m-\frac{1}{2}}(\chi)$--functions are Legendre
functions of the second kind and half--integer degree. We evaluate
them using a modified\footnote{It is modified in the sense that the
  limit on how close to one the argument $\chi$ can be is
  changed. This was done in order to get sufficient numerical
  precision.} version of software \textsf{DTORH1.f} described
in\cite{Segura}.

Using Eqs.~(\ref{wavefunctionexpansion_eq}) and~(\ref{multipole_eq}), we
can write the electron--electron interaction matrix element as
\begin{eqnarray}
\label{ee_interaction_BIG_eq}
& & \langle ab| \frac{1}{\hat{r}_{12}}| cd \rangle = \nonumber \\
& & \frac{e^2}{4\pi\epsilon_r\epsilon_0} 
\langle u_{a}(r_i)u_{b}(r_j)|
\frac{Q_{m-\frac{1}{2}}(\chi)}{\pi\sqrt{r_ir_j}}
| u_{c}(r_i)u_{d}(r_j) \rangle \nonumber \\
& \times &
\langle e^{im_{\ell}^a\phi_i}e^{im_{\ell}^b\phi_j}|
\sum_{m=-\infty}^{\infty} e^{im(\phi_i-\phi_j)}
|e^{im_{\ell}^c\phi_i}e^{im_{\ell}^d\phi_j}\rangle \nonumber \\
& \times &
\langle m_s^a|m_s^c\rangle
\langle m_s^b|m_s^d\rangle.
\end{eqnarray}

Note that the angular ($\phi$) integration in
(\ref{ee_interaction_BIG_eq}) yields a non-zero result only if
$m=m_{\ell}^a - m_{\ell}^c = m_{\ell}^d - m_{\ell}^b$. This determines
the degree $m$ of the Legendre--function in the radial part of
Eq.~(\ref{ee_interaction_BIG_eq}). It is also clear from
Eq.~(\ref{ee_interaction_BIG_eq}) that the electron--electron matrix
element equals zero if orbital  $a$ and $c$ or orbitals $b$ and $d$ have
different spin directions.

\subsection{Alternative single particle potentials}
If $V$ in Eq.~(\ref{perturb}) has to account for the full
electron-electron interaction, as it does when the single particle
Hamiltonian just includes the external confinement potential,
Eq.~(\ref{harmpot}), it might be difficult to obtain convergence of the
iterative solutions of Eq.~(\ref{eq:bloch}). At least this is expected
for weak external confinement and for many-electron dots. A remedy is
then to start from a Hamiltonian $H_0$ that already includes the bulk
of the electron-electron interaction. Many choices are possible here
and we have investigated two of these.  For a single Slater
determinant the Hartree-Fock approximation is known to minimize the
energy.  In this approximation, each electron moves in the average
potential from the other electrons.  The one-particle potential, including the
external confinement, is then 
\begin{eqnarray}
u^{HF}_{ji} & =  & \langle B_j| \hat{u}_c(r) |B_i \rangle \nonumber \\ & +  &
\sum_{a\le N} \langle B_j a|\frac{1}{\hat{r}_{12}}|B_i a\rangle-\langle B_ja|\frac{1}{\hat{r}_{12}}|aB_i\rangle.
\label{uHF}
\end{eqnarray}
The last term gives the (non-local) exchange interaction.  One
complication with Eq.~(\ref{uHF}) is that for a situation where not
all electron spins are paired, electrons with the same quantum numbers
$n$ and $m_{\ell}$ but with different spin directions will
experience different potentials. As a consequence, the total spin ${\bf
  S}^2 = \left( \sum_i {\bf s}_i\right)^2$, does not commute with the
Hartree-Fock Hamiltonian. In spite of the fact that the full
Hamiltonian, Eq.(\ref{perturb}), still commutes with ${\bf S}^2$, this
property might lead to complications, see Ref.\cite{waltersson:07} for
more details, and it might be more practical to use a starting point
where the exchange interaction is approximated with a local
potential. We adopt for this a traditional Local Density Approximation (LDA)
with a variable amount of exchange
\begin{eqnarray}
u^{LDA}_{ji} 
& =  & \langle B_j| \hat{u}_c(r)  |B_i \rangle  +
\sum_{a\le N} \langle B_j a|\frac{1}{\hat{r}_{12}}|B_i a\rangle \nonumber  \\ & -  &
  \eta \langle B_j| 4 a_B^* \sqrt{\frac{2 \rho(r)}{\pi}} | B_i\rangle,
\label{uLDA}
\end{eqnarray}
 where $\rho(r)$ is the radial electron density and $\eta$ is the so
 called Slaters exchange parameter, which often is set to one.  In
 Section~\ref{sec:results} we present results with $\eta = 1.0$ and $\eta
 =1.4$.

\subsection{Many-Body treatment}
Equipped with a finite representation of the $Q$ space it is possible
to construct the $S_n$-operators, and thus also the wave operator
$\Omega$. We now use the Coupled Cluster Singles and Doubles
truncation of the possible excitations, i.e. only the terms in the
boxes in Eq.~(\ref{scluster}) are kept. We start from
Eq.~(\ref{eq:soperator}) and note that for a model space built from a
single Slater determinant, the $\chi PV\Omega P$-term is fully
canceled by the unlinked diagrams from the $QV\Omega P$-term. Only the
$QV\Omega P$-term remains thus on the right-hand side of
Eq.~(\ref{eq:soperator}).  Starting with $\Omega^{(1)}=1$ and
$\chi^{(1)}=0$, we can write the recursion relation for the the
$S_1$-amplitudes as
\begin{multline}
\label{S1}
\langle \alpha_a^r | S_1 | \alpha \rangle^{i+1} = \frac{1}{ \epsilon_a - \epsilon_r } \langle \alpha^r | V_1 + VS_1 + VS_2 \\ 
+\frac{1}{2!} V\{ S_1^2\} + V_2 \{ S_1 S_2\} + \frac{1}{3!} V_2 \{ S_1^3 \} | \alpha \rangle^{i} 
\end{multline}
and for the $S_2$-amplitudes\begin{multline}
\label{S2}
\langle  \alpha_{ab}^{rs} | S_2 | \alpha \rangle^{i+1} = 
\frac{1}{ \epsilon_a +  \epsilon_b- \epsilon_r - \epsilon_s} \langle \alpha_{ab}^{rs} |  
V_2 + V_2 S_1 + VS_2 \\ 
+ \frac{1}{2!} V_2 \{ S_1^2\} + V \{ S_1 S_2\} + \frac{1}{3!} V_2 \{ S_1^3 \} + 
\frac{1}{2!} V_2 \{ S_2^2 \} \\
+ \frac{1}{2!} V_2 \{ S_1^2 S_2 \} + \frac{1}{4!} V_2 \{ S_1^4 \} | \alpha \rangle^{i}, 
\end{multline}
where only connected contributions should be kept on the right-hand
side.  Here $V=V_1+V_2$ is the total perturbation, $V_1$ is the part
of the perturbation that can be written as a one-particle operator and
$V_2$ is the part of the perturbation that can be written as a two
particle operator. The index $i$ denotes the iteration number. It is
related, but not equal, to the order in the perturbation
expansion. The quoted figures in section \ref{sec:results} are always
self-consistent with respect to Eqs.(\ref{S1}-\ref{S2}).  Note that,
e.g. the single excitation cluster, $S_1$ (\ref{S1}), is built from up
single, double and triple excitations. As an example we note that the
included triples are those that can be be written as disconnected
singles connected by a perturbation $V_2$ (the last term on the second
line of Eq.~(\ref{S1})), as well as combinations of singles and
doubles connected by $V_2$ (second term on the last line of
Eq.(\ref{S1})). These are so called \emph{intermediate} triple
excitations. In a similar way also intermediate triples and quadruples
contribute to $S_2$.

Finally, it is appropriate to comment on the difference between the
two-dimensional many-body procedure and the more studied
three-dimensional case, especially with respect to the angular
integration.  The angular momentum algebra is considerably simplified
in two dimensions compared to three dimensions.  In two dimensions an
orbital is defined by only three quantum numbers.  With polar
coordinates these are the radial quantum number, $n$, the angular
quantum number, $m_{\ell}$, and the spin direction, $m_s$. The radial
functions $u_{n m_{\ell} m_s}(r)$,
Eq.(\ref{wavefunctionexpansion_eq}), depend on two of these quantum
numbers, $n, m_{\ell}$, while an additional dependence on $m_s$ only
arises in case an external magnetic field is applied to the dot. In
three dimensions the desired total angular momentum has to be
constructed through a linear combination of the different magnetic
components of the orbitals. In spite of the advanced formalisms
(e.g. Racah algebra) developed in order to avoid explicit summation
over magnetic substates, the angular integration often gets rather
cumbersome, at least for general open shell configurations. In two
dimensions there is only place for one particle in each spatial
orbital and any state with maximized total spin can be treated as a
closed shell configuration is handled in there dimensions.


\section{Results}\label{sec:results}
In our numerical studies we use $m^*=0.067m_e$ and $\epsilon_r=12.4$
corresponding to the bulk value in GaAs.
\subsection{Validation}

\begin{table*}[ht!]
\caption{\label{tab:CIcompare} Comparison between the Coupled Cluster
  Singles and Doubles method starting from the pure one-electron basis
  (this work) and Full Configuration Interaction according to the
  software developed by Kvaal~\cite{kvaal:09} and according to Rontani
  {\em et al.}~\cite{Rontani2006}. For truncation of the basis set the so
  called shell truncation parameter $R = 2n + |m_l|$ is used. Energies
  are given in units of $\hbar\omega$ and the number of confined
  electrons is $2$-$6$ and $8$.  $\lambda$ is defined in Eq.(\ref{eq:lambda}).}
\begin{ruledtabular}
\begin{tabular}{cc|cc|c|ccc} 
    &                  &                      &                     &          & \underline{CCSD}& \multicolumn{2}{c}{\underline{\hspace{40pt} FCI \hspace{40pt} }}\\
    &$|2S ~M_L\rangle$ & Confinement Strength & $\hbar\omega (meV)$ &Basis set & This work & Kvaal & Rontani {\em et al.}\\
\hline
N=2 &$|0 0\rangle$   &  $\lambda$ = 1.0& 11.85720 &R=5 & 3.013625 &  3.013626\footnotemark[1] & \\
    &                &                 &          &R=6 & 3.011019 &  3.011020\footnotemark[1] & \\
    &                &                 &          &R=7 & 3.009234 &  3.009236\footnotemark[1] & \\
    &                &                 &          &                  &  &   & \\
    &                &  $\lambda$ = 2.0& 2.964301 &R=5 & 3.733597&  3.733598\footnotemark[1] & 3.7338\\
    &                &                 &          &R=6 & 3.731057&  3.731057\footnotemark[1] & 3.7312\\
    &                &                 &          &R=7 & 3.729323&  3.729324\footnotemark[1] & 3.7295\\
    &$|2 1\rangle$   &  $\lambda$ = 2.0&          &R=5 & 4.143592&  4.143592\footnotemark[1] & 4.1437 \\
    &                &                 &          &R=6 & 4.142946&  4.142946\footnotemark[1] & 4.1431 \\
    &                &                 &          &R=7 & 4.142581&  4.142581\footnotemark[1] & 4.1427 \\
    &$|0 0\rangle$   &  $\lambda$ =6.0& 0.3293668 &R=5 & 5.784651&           & 5.7850\\
    &$|0 0\rangle$   &  $\lambda$ =8.0& 0.1852688 &R=5 & 6.618102&           & 6.6185\\
    &                &                &           &R=6 & 6.618091&           & 6.6185\\
    &                &                &           &R=7 & 6.618089&           & 6.6185\\
\hline                                              
N=3 &$|1 1\rangle$   &  $\lambda$ =1.0& 11.85720 &R=6 & 6.37600  & 6.374293\footnotemark[2] &       \\
    &                &                &          &R=7 & 6.37293  & 6.371059\footnotemark[2] &       \\
    &                &                &          &R=8 & 6.37069  & 6.368708\footnotemark[2] &       \\
    &                &                &          &R=10& 6.36773  & 6.365615\footnotemark[2] &       \\
    &                &  $\lambda$ =2.0& 2.964301 &R=5 & 8.18306  & 8.175035\footnotemark[2] & 8.1755\\
    &                &                &          &R=6 & 8.17896  & 8.169913\footnotemark[2] &       \\
    &                &                &          &R=7 & 8.17635  & 8.166708\footnotemark[2] & 8.1671\\
\hline                                              
N=4 &$|2 0\rangle$   &  $\lambda$ =2.0& 2.964301 &R=7 & 13.635   &                & 13.626\\
\hline                                              
N=5 &$|1 1\rangle$   &  $\lambda$ =2.0& 2.964301 &R=5 & 20.3697  &                & 20.36\\
    &                &                &          &R=6 & 20.3554  &                & 20.34\\
    &                &                &          &R=7 & 20.3467  &                & 20.33\\
\hline                                                  
N=6 &$|0 0\rangle$   &  $\lambda$ =2.0& 2.964301 &R=5 & 28.0161  & 28.0330\footnotemark[3]& 28.03\\
    &                &                &          &R=6 & 27.9912  &                & \\
    &                &                &          &R=7 & 27.9751  &                & 27.98\\
    &                &                &          &R=10& 27.9529  &                & \\
    &                &                &          &R=15& 27.9390  &                & \\
\hline                                                  
N=8 &$|2 0\rangle$   &  $\lambda$ =2.0& 2.964301 &R=5 & 47.13801 &                & 47.14 \\
    &                &                &          &R=10& 46.70369 &                & \\
    &                &                &          &R=15& 46.67960 &                & \\
\end{tabular}
\footnotetext[1] {Simen Kvaal\cite{kvaal:09}}
\footnotetext[2] {Simen Kvaal\cite{kvaal:private},
  obtained with the  software in Ref.~\cite{kvaal:09}}
\footnotetext[3] {Patrick Merlot~\cite{merlot_ex_jobb} using the software in Ref.~\cite{kvaal:09}.}
\end{ruledtabular}
\end{table*}

Table~\ref{tab:CIcompare} compares the present results with those
obtained by Full Configuration Interaction,
Refs.~\cite{kvaal:09,merlot_ex_jobb,Rontani2006}, for $2$-$6$ and $8$
electrons when starting from the non-interacting basis.  The purpose
of Table~\ref{tab:CIcompare} is first to compare with calculations
which include exactly the same physical effects.  Such a comparison
can in principle only be done for two electrons due to the truncation
of $S_3$-clusters and beyond in the CCSD method. A second purpose is
to compare the accuracy and basis set convergence for more than two
electrons and for confinement strengths close to the region of
interest.

The key parameter here is the strength of the electron-electron
interaction relative the confinement provided by the harmonic
oscillator potential, which can be quantified, as in
Table~\ref{tab:CIcompare}, by the length parameter $\lambda$,
\begin{equation}
\label{eq:lambda}
\lambda = \frac{\hbar}{\omega m^*} \frac{m^* e^2}{4\pi \varepsilon_0 \varepsilon_r \hbar^2} =
\frac{\hbar}{\omega m^*} \frac{1}{a_0^*},
\end{equation}
where $m^*$ is the effective mass, $\epsilon_r$ is the relative
dielectric constant, and $a_0^*$ is the effective Bohr radius.  We
recall here that $\sqrt{\frac{\hbar}{\omega m^*}}$ is the harmonic
oscillator length unit.  Larger $\lambda$-values correspond to a
weaker confinement and an increased relative importance of the
electron-electron interaction. For Table~\ref{tab:CIcompare} we
(mainly) choose $\lambda=2$ corresponding to $\hbar\omega\approx2.964$ meV
(for GaAs parameters) since the next available CI-results are either
much stronger or much weaker than this.

%
%
%

For two electrons, $N=2$, both the CCSD and the FCI method take all
electron-electron effects into account and the accuracy is thus only
limited by the size of the basis set and the numerical procedure.
When comparing the $N=2$ results produced with identical basis sets in
Table~\ref{tab:CIcompare}, we note that our values differ from those
produced by Kvaal~\cite{kvaal:09} at most in the seventh digit,
while those by Rontani {\em et al.}~\cite{Rontani2006} differ in the
fifth digit. The computer code developed by Kvaal~\cite{kvaal:09} is
bench-marked to machine precision with exact results and it is
reasonable to believe that its numerical accuracy is the highest.  The
leading numerical errors in the present implementation are due to the
precision in the $Q_{m-1/2}$-functions produced by \textsf{DTORH1.f}
described in\cite{Segura}, and their integration in
Eq.(\ref{ee_interaction_BIG_eq}).  For more than two electrons these
numerical errors are much smaller than the errors introduced through
truncations, e.g. of basis sets or of $S_3$ clusters and beyond, and
are of no significance. For the the remaining part of
Table~\ref{tab:CIcompare} we restrict the display of our results to
six digits.

Since coupled-cluster is an iterative and perturbative method,
convergence is never guaranteed.
We note  that for $N=2$ in Table~\ref{tab:CIcompare}, convergence
is still obtained for potential strengths as low as $\lambda=8$ even though
the full electron-electron interaction is here taken as the
perturbation. We emphasize that this constitutes a truly
non-perturbative case; the total energy is almost seven times as large
as the strength of the confining potential.

Continuing to $N>2$, we conclude that 
the  largest relative error, calculated as the percentage of
the total energy, arises for $N=3$. Still the deviation is never
larger than $\sim 1.5\times 10^{-2}$ in units of $\hbar\omega$ or
$\sim 0.1$ percent of the total energy.
For $6$ and $8$ electrons we also increased the basis set
significantly beyond what so far has been feasible with FCI.  It is
clear, at least for $N\ge 6$ and the potential strengths studied here,
that the error made by the truncation of the Coupled Cluster expansion
to include only the $S_1$ and $S_2$ cluster operators, see
Eq.(\ref{scluster}), is far smaller than the error made by truncating
the basis set in the CI-calculations.



Table~\ref{tab:CIcompare}  shows only
results
obtained with the whole electron-electron interaction as a
perturbation. The reason for this is the wish to be able to compare
with CI-calculations.  These typically use a non-interacting basis
set, which further is severely truncated. With this starting point,
convergence of the coupled-cluster expansion  for weaker confinements than $\lambda=2$, for $N>2$ can be
problematic. However, convergence can be obtained for a much wider range of
confinement strengths with a better starting point: e.g. Hartree-Fock
or Local density, but fair comparison can then only be made with
converged results. 

\begin{table*}[ht!]
\caption{\label{tab:3elcomparison} The importance of $S_3$ clusters
  and beyond. The present Coupled Cluster Singles and Doubles results
  are compared to Full Configuration
  Interaction\cite{kvaal:09,kvaal:private,merlot_ex_jobb} results
  obtained with the same basis sets. For three electrons the basis is
  for both methods truncated at $R=2n+|m_l|=7$, and with six electrons
  it is truncated at $R=5$.  These basis sets are {\em not} saturated,
  but the comparison unveils the level at which contributions beyond
  CCSD contribute, as function of the confinement strength.  The
  values in parenthesis are the differences to the corresponding Full
  CI-value. Energies are given in units of $\hbar\omega$. }
\begin{ruledtabular}
\begin{tabular}{cc |cc|ccc}
&$|2S~M_L\rangle$& $\lambda$ & $\hbar\omega (meV)$ & CCSD (present) & 
Full CI\\
\hline
N=3&$|1 1\rangle$   & 0.5    &   47.42881   & 5.28660 (+0.00019)& 5.28640\\
   &                & 0.75   &   21.07947   & 5.85048 (+0.00077)& 5.84971\\
   &                & 1.0    &   11.85720   & 6.37292 (+0.00187)& 6.37106\\
   &                & 1.5    &   5.269868   & 7.32169 (+0.00539)& 7.31630\\
   &                & 2.0    &   2.964300   & 8.17635 (+0.00965)& 8.16670\\
   &                & 4.0    &   0.7410752  & \textrm{diverges}& 11.0425\\
   &                &        &              &                   &        \\
   &$|3 0\rangle$   & 0.5    &   47.42881   &5.90815 (+5$\times 10^{-6}$) &5.90814\\
   &                & 0.75   &   21.07947   &6.34019 (+0.00002)         &6.34017\\
   &                & 1.0    &   11.85720   &6.75908 (+0.00006)         &6.75903\\
   &                & 1.5    &   5.269868   &7.56147 (+0.00020)        &7.56128\\
   &                & 4.0    &   0.7410752  &11.0514 (~-0.00120)        &11.0526\\
\hline
N=6 &$|0 0\rangle$  &  0.1   & 1185.720    & 11.1979 (+8$\times 10^{-7}$) & 11.1979\\
    &               &  0.5   & 47.42881    & 15.5624 (+0.00062) & 15.5618\\
    &               &  1.0   & 11.85720    & 20.2609 (+0.00371) & 20.2572\\
    &               &  2.0   & 2.964301    & 28.0161 (~-0.01687) & 28.0330\\
\end{tabular}
\end{ruledtabular}
\end{table*}

\begin{figure}[ptb]
\includegraphics[width=\columnwidth]{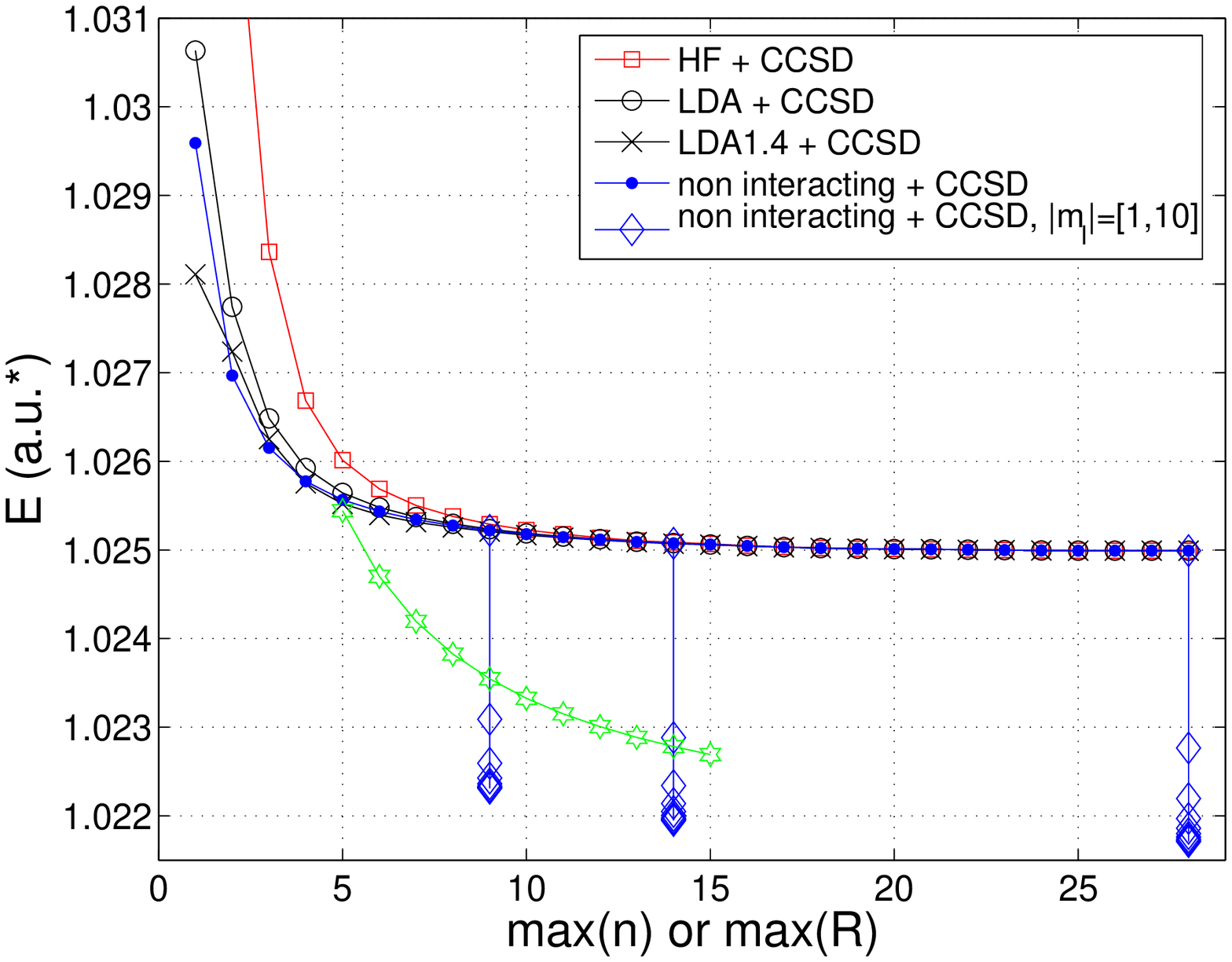}
\caption{ The figure shows the $n$-convergence of the two-particle ground state energy 
for a dot with a confining potential corresponding to $\hbar\omega=3.32$meV.
The  $n$-convergence is studied for a variety of starting points  in  the CCSD-calculation.
The maximum $n$- value is varied from $1$ to $28$, while $|m_l|\le 1$
for a Hartree-Fock starting point (red), different Local Density
starting points (black), and a non-interacting starting point
(blue with solid circles).
  The vertical lines with diamonds for max$(n)=9,14$ and $28$ show
  the $|m_l|$-convergence with max$(|m_l|)=[1,10]$. For reference we
  note that the energy using second order perturbation theory on top
  of Hartree Fock is $1.045$ a.u.* when using the basis size
  $(n,|m_l|)\le (28,1)$. The (green) curve with hexagram markers shows
  the $R=2n+|m_l|$-convergence when starting from the non-interacting
  basis set. \label{twoel_diff_start_points}}
\end{figure}


A strict limitation with the CCSD approach is the neglect of true
triples, $S_3$ clusters, and beyond. For sufficiently weak
confinements this approximation will dominate the
error. Table~\ref{tab:3elcomparison} shows a comparison between the
CCSD and FCI methods for three and six electrons and for a large range
of confinement strengths. The purpose is here to establish how
important the limitation to $S_1$ and $S_2$ clusters is. Confinement
strengths as weak as possible, but still leading to a converged
coupled-cluster expansion with a non-interacting basis have been used.
We note that for $\lambda \le 1$ the CCSD-method yields results
accurate enough for most practical purposes. For $\lambda = 2$ the
error due to the neglected effects in the CCSD-method is still so
small that (keeping table \ref{tab:CIcompare} in mind) the possibility
to use larger basis sets than in a CI calculation well compensates for
the lack of triples and beyond. For $N=3$ and $\lambda=4$ the CCSD did
not converge for the $|2S M_L 1\rangle = |1 1\rangle$ ground state.
For this weak confinement the first excited state is still
reproduced well. Generally we see that this first excited state, which
is not as localized as the ground state, is reproduced better than the
ground state throughout the list of confinements strengths in Table~\ref{tab:3elcomparison} . Intuitively this makes sense; true triple
excitations (and all many-body effects) should be 
relatively more important for more 
 localized states.

\subsection{Convergence of the basis and the use of different starting points}

 Fig.~\ref{twoel_diff_start_points} depicts the basis set convergence
 for the $N=2$ ground state using different starting points. The
 potential strength $\hbar\omega=3.32$~meV, corresponding to $\lambda
 \approx 1.89$, is chosen to enable comparison with the Quantum Monte
 Carlo results from Ref.~\cite{montecarlo:err:03}, where it is argued
 that this value is close to the actual values in the experiment by
 Tarucha {\em et al.}~\cite{Tarucha}.
 Fig.~\ref{twoel_diff_start_points} shows that the basis sets begin to
 saturate
around max$(n)=10$ 
 and for the angular quantum number 
 they saturate to
 the same extent at approximately $|m_l|=
4$. The shell truncation parameter $R=2n+|m_l|$, used in
most of the previous CI-calculations, is thus here ($N=2$ and $\hbar\omega
=3.32$~meV) clearly overemphasizing the 
need for high
$|m_l|$-values in the basis set, on the expense of $n$-values.
%
To put it in other words; it is apparent from the figure
that a basis cut of $(n,|m_l|)\le (10,4)$ is a better choice than
e.g. $R=2n+|m_l|\le 10$. 
For weaker confinements and more confined particles the behavior is
probably different; one expects then the high $|m_l|$ basis functions
to be relatively more important.

 Moreover, Fig.~\ref{twoel_diff_start_points} shows that for small $n$
 the different starting points give quite a spread in the energy but
 for n$>10$ the results are virtually independent of the starting
 point. Again this demonstrates that for the comparison with
 CI-results from Refs.~\cite{kvaal:09,merlot_ex_jobb,Rontani2006},
 which all are limited to very few $n$:s, we are limited to
 the use of the non-interacting starting point.  This is unfortunate
 since the CCSD method can handle much weaker confinement strengths
 with a Hartree-Fock or a Local Density Starting point.

It is easy to be mislead by Fig.~\ref{twoel_diff_start_points} and
draw the conclusion that there is no need for other starting points
than the pure harmonic oscillator basis and that Hartree-Fock is the
worst of the tested starting points. However, the number of needed
{\em iterations} in Eq.(\ref{S1} - \ref{S2})($\sim$ orders in the
perturbation expansion) to obtain self-consistency can differ
significantly between the starting points, with Hartree-Fock often
being the fastest of them all. For example, with a basis size of
$(n,|m_l|)\le (28,1)$ and including up to second order corrections to
the energy, the pure harmonic oscillator starting point yields
$0.9223$~a.u.*, 
while Hartree-Fock gives $1.0449$~a.u.* and the fully converged result
is $1.0250$~a.u.* (for both starting points).  That is, 
with the Hartree-Fock basis we start much closer to the fully correlated situation.

\subsubsection{Extrapolation and convergence of the basis}

Even though much larger basis sets can be used with CCSD than with CI,
the calculation time still grows with the number of particles and with
the size of the basis set.  Extrapolation of the results to infinitely
large basis sets is then often an efficient strategy.  Many elaborate
strategies can be envisaged here, e.g. adjusting the size of the basis
set during the iterations. One can for example obtain convergence in a
limited basis and then systematically increase it, or one can filter
the contributions and only keep those that are estimated to contribute
over a certain level. Here we restrict ourselves to a brief discussion
of the potential gain of extrapolation.

\begin{table}[ht!]
\caption{\label{tab:extrapolation2el_ml} The $|m_l|$-convergence
  (using all $n$, in this case $n\le 29$) of the two electron ground
  state energy given in a.u.*. The extrapolated values are found
  according to the procedure described in the text. The extrapolated
  values agree well with the values 1.02164(1) and 1.02165(1) by
  Pederiva et al~\cite{montecarlo:err:03} obtained through Variation
  Monte Carlo and Diffusion Monte Carlo methods respectively. For each
  extrapolated value three points were used in the linear fit.}
\begin{ruledtabular}
\begin{tabular}{ccccc} 
 $|m_l|$ &  E($|m_l|$) &  $|m_l|\rightarrow\infty$ & Upper bound& Lower bound\\
\hline
 1  &   1.024993 &  -  & -  & - \\
 2  &   1.022767 &  -  & -  & -  \\
 3  &   1.022196 &    1.021703  & 1.022196  & 1.021210 \\
 4  &   1.021971 &    1.021667  & 1.021971  & 1.021363 \\
 5  &   1.021861 &    1.021661  & 1.021861  & 1.021461 \\
 6  &   1.021799 &    1.021660  & 1.021800  & 1.021520 \\
 7  &   1.021762 &    1.021660  & 1.021762  & 1.021558\\ 
 8  &   1.021737 &    1.021661  & 1.021738  & 1.021584\\
 9  &   1.021721 &    1.021661  & 1.021721  & 1.021601\\
 10 &   1.021709 &    1.021662  & 1.021709  & 1.021614\\
\end{tabular}
\end{ruledtabular}
\end{table}

\begin{table*}[ht!]
\caption{\label{tab:extrapolation2el_R} The $R=2n+|m_l|$-convergence
  for the $2 - 8$ electron ground states  for
  a confining potential corresponding to $\hbar \omega = 3.32$~meV. The results are  
  given in a.u.*}
\begin{ruledtabular}
\begin{tabular}{cccccccc} 
 $R$ &  2e$^-$ &  3e$^-$ & 4e$^-$  & 5e$^-$   & 6e$^-$   & 7e$^-$   &  8e$^-$ \\ 
\hline
 5  & 1.02544  & 2.24045 & 3.72652 & 5.55317  & 7.62640  & 10.0913& 12.8065  \\ 
 6  & 1.02470  & 2.23924 & 3.72439 & 5.54890  & 7.61942  & 10.0624& 12.7284  \\ 
 7  & 1.02420  & 2.23846 & 3.72210 & 5.54629  & 7.61496  & 10.0541& 12.7132  \\ 
 8  & 1.02383  & 2.23791 & 3.72207 & 5.54454  & 7.61207  & 10.0498& 12.7067  \\ 
 9  & 1.02355  & 2.23750 & 3.72140 & 5.54327  & 7.61000  & 10.0469& 12.7027  \\ 
 10 & 1.02333  & 2.23719 & 3.72089 & 5.54234  & 7.60846  & 10.0448& 12.7000  \\ 
 11 & 1.02315  & 2.23694 & 3.72050 & 5.54161  & 7.60728  & 10.0431& 12.6979  \\ 
 12 & 1.02301  & 2.23674 & 3.72018 & 5.54102  & 7.60633  & 10.0419& 12.6962  \\ 
 13 & 1.02289  & 2.23657 & 3.71991 & 5.54054  & 7.60557  & 10.0408& 12.6949  \\ 
 14 & 1.02278  & 2.23643 & 3.71969 & 5.54014  & 7.60493  & 10.0400& 12.6939  \\ 
 15 & 1.02269  & 2.23631 & 3.71950 & 5.53979  & 7.60438  & 10.0393& 12.6930  \\ 
 \hline
 QMC\footnotemark[1] 
 & 1.02165(1) & 2.2395(1) & 3.7194(1)& 5.5448(1)& 7.6104(1) & 10.0499(1)& 12.7087(1) \\ 
  QMC\footnotemark[2]
 & 1.02164(1) & 2.2339(1) & 3.7145(1)& 5.5338(1)& 7.6001(1) & 10.0342(1)& 12.6900(1) \\ 
\end{tabular}
\end{ruledtabular}
\footnotetext[1] {Variation Monte Carlo, Pederiva {\em et al.}\cite{montecarlo:err:03}}\\
\footnotetext[2] {Diffusion Monte Carlo, Pederiva {\em et al.}\cite{montecarlo:err:03}}
\end{table*}

 A first example is shown in Table~\ref{tab:extrapolation2el_ml},
 where the full radial basis is used to investigate the $m_{\ell}$
 expansion. The extrapolated values are obtained through a linear
 regression fit assuming the relation
\begin{equation}
\ln \left[ E(|m_{\ell}|+1)-E(|m_{\ell}|) \right] = K \ln(|m_{\ell}|+1) + C, 
\end{equation}
where $E(|m_{\ell}|)$ is the energy with the one particle basis cut at
$\textrm{max}(|m_{\ell}|) = |m_{\ell}|$ and $K$ and $C$ are the
constants we find from the fit. The fits were made with three
consecutive differences at a time, which give the list of predictions
displayed in the third column in
Table~\ref{tab:extrapolation2el_ml}. It is clear that this procedure
can improve the results substantially, especially when a rather low
maximum $|m_{\ell}|$ is used. The upper (lower) bound decreases
(increases) monotonically as $|m_l|$ increases and the extrapolated
values stabilize around $1.02166$ a.u.*. This result agrees well with
those obtained by Pederiva {\em et al}~\cite{montecarlo:err:03},
$1.02164(1)$ and $1.02165(1)$ a.u.*, through variational and diffusion
quantum Monte Carlo methods respectively.

Table~\ref{tab:extrapolation2el_R} shows the convergence of the ground
state energies for two to eight confined electrons as functions of
$R=2n+|m_l|$.  This truncation scheme is a common choice in numerical
studies\cite{Rontani2006,kvaal:09}, and the motivation to study
convergence as a function of this parameter is that the energy levels
of a non-interacting two-electron dot are given by
$\epsilon_{nm_{\ell}} = (2n + |m_{\ell}| + 1) \hbar \omega$. The
confining potential corresponds to $\hbar \omega = 3.32$~meV, a
strength that previously has been studied with Quantum Monte Carlo
Methods\cite{montecarlo:err:03}, which are shown in the Table for
comparison.  A comparison with the two-electron results from
Table~\ref{tab:extrapolation2el_ml} indicates though that the radial
convergence is substantially slower than the angular convergence, at
least for this confinement strength. For more than two particles the
angular convergence slows down due to the mutual repulsion felt by the
electrons and the 
resulting spread of the total wave function. This makes $R$ a
reasonable parameter for the truncation of the basis.  We emphasize
that we have used considerably higher $R$ and/or more confined
electrons than used in the available CI
calculations\cite{kvaal:09,Rontani2006}. The error relative to the
Diffusion Monte Carlo method (expected to be the more accurate of the
two\cite{montecarlo:err:03}) is $\sim 10^{-3}$ and on the same level
or below as the difference between the two different Monte Carlo
methods. For most purposes so far this would, we believe, be a good
enough convergence. We stress that the complete series of $R=[5,15]$
for $N=8$ still did not take more than $\sim 24$~hours to compute on a
standard desktop machine, which implies that further converged results
can be obtained if necessary.

\begin{table}[ht!]
\caption{\label{tab:extrapolation2el_alt} The ground state energy of
  the $\hbar\omega = 3.32$~meV two-electron dot obtained with an
  alternative extrapolation scheme: Aitken's $\delta^2$-process is
  used on the $n=7,8$ and $9$ values for respective $|m_l|$ to
  accelerate the $n$-convergence. Subsequently the
  $|m_l|\rightarrow\infty$ extrapolation is done as described in the
  text. The results are given in a.u.$^*$.}
\begin{ruledtabular}
\begin{tabular}{ccccc} 
 $|m_l|$ &  E($|m_l|$) &  $|m_l|\rightarrow\infty$ & Upper bound& Lower bound\\
\hline
 1  & 1.025029\footnotemark[1]   &  -  & -  & - \\
 2  & 1.022809\footnotemark[1]   &  -  & -  & - \\
 3  & 1.022232\footnotemark[1]   &  -  & -  & - \\
 4  & 1.022001\footnotemark[1]   & 1.021697   & 1.021731  & 1.021658\\
 5  & 1.021893\footnotemark[1]   & 1.021703   & 1.021719  & 1.021687\\
 6  & 1.021839\footnotemark[1]   & 1.021734   & 1.021737  & 1.021697\\
\end{tabular}
\end{ruledtabular}
\footnotetext[1] {Obtained by Aitken's $\delta^2$-process for $n=7-9$.}
\end{table}

However, it seems to be very difficult to improve the $R$-cut results
through extrapolation.  All our attempts so far have resulted in quite
unreliable predictions. For example, the extrapolated values did not
stabilize, but showed a monotonously decreasing behavior. Instead we
have investigated a third possible truncation scheme. In this scheme
we have first used the Aitken's $\delta^2$-process, see
e.g. Ref.~\cite{NumRec}, to speed up the convergence in $n$.  After
that we extrapolate to infinity in $|m_l|$ as described at the
beginning of this subsection in connection with
Table~\ref{tab:extrapolation2el_ml}. The results for the two-electron
dot are displayed in Table~\ref{tab:extrapolation2el_alt}, where
values for $n=7,8$ and $9$ for each $|m_l|-$value were used to perform
Aitken's convergence acceleration. In contrast to the attempted
$R$-extrapolation, the extrapolated values do now stabilize, and the
obtained value $1.0217$ a.u.* is very close to the result obtained in
Table~\ref{tab:extrapolation2el_ml} using a much larger basis set. The
usage of this or similar extrapolation schemes for more than two
confined particles is an interesting topic for future studies.



\section{Conclusions}

In conclusion, by comparison with results obtained with Full
Configuration Interaction and two different Quantum Monte Carlo
methods the Coupled Cluster Singles and Doubles approach is shown to
be a very powerful method for two to eight electrons confined in a
two-dimensional harmonic oscillator potential. For $\lambda\le 1$
($\hbar\omega \approx 11.857$meV for GaAs material parameters) the
results are for practical purposes exact when comparing with FCI and
for $\lambda=2$ ($\hbar\omega \approx 2.964$meV for GaAs material
parameters) the error is still never larger than $\sim1.5\times
10^{-2}$ in units of $\hbar\omega$. For $N\ge 6$ the possibility to
use much larger basis sets than in FCI-calculations is shown to be of
uttermost importance. The errors introduced by truncating the basis
sets in FCI-calculations are in many cases much larger than the error
made by truncating some of the triple and quadruple excitations as
done in CCSD. Moreover, when comparing with a Diffusion Monte Carlo
study, for a potential strength close to what is estimated from the
experiment by Tarucha et al.\cite{Tarucha}, the errors in the two to
eight electron ground states are shown to be on the same level or less
than the differences between Variational and Diffusion Monte Carlo
results. This gives promise that the method, in future studies, can be
used for the extraction of reliable information from experiments.

\begin{acknowledgments}
Financial support from the Swedish Research Council(VR) and from the
G{\"o}ran Gustafsson Foundation is gratefully acknowledged. We also
want to thank  Simen Kvaal who provided us with many
Configuration Interaction results for reference.
\end{acknowledgments}


\end{document}